\documentclass[sigconf]{acmart}

\usepackage{booktabs} 

\usepackage{balance}       
\usepackage{graphics}      
\usepackage[T1]{fontenc}   
\usepackage{color}
\usepackage{float}
\usepackage{tabularx}
\usepackage{relsize}
\usepackage{url}
\usepackage{enumitem}

\def\plaintitle{A Maturity Assessment Framework for \\ Conversational AI Development Platforms}

\newcommand{\compactlist}{\setlength{\itemsep}{0pt} \setlength{\parskip}{0pt}}

\copyrightyear{2021}
\acmYear{2021}
\setcopyright{acmlicensed}\acmConference[SAC '21]{The 36th ACM/SIGAPP Symposium on Applied Computing}{March 22--26, 2021}{Virtual Event, Republic of Korea}
\acmBooktitle{The 36th ACM/SIGAPP Symposium on Applied Computing (SAC '21), March 22--26, 2021, Virtual Event, Republic of Korea}
\acmPrice{15.00}
\acmDOI{10.1145/3412841.3442046}
\acmISBN{978-1-4503-8104-8/21/03}

\begin{document}
\title{\plaintitle{}}

\author{Johan Aronsson, Philip Lu}
\affiliation{%
  \institution{Chalmers | University of Gothenburg, Sweden}
}
\email{convaistudy@easelab.org}

\author{Daniel Strüber}
\affiliation{%
  \institution{Radboud University, Nijmegen, Netherlands}
}
\email{d.strueber@cs.ru.nl}

\author{Thorsten Berger}
\affiliation{%
  \institution{
	Ruhr University Bochum, Germany;
Chalmers | University of Gothenburg, Sweden}
}
\email{thorsten.berger@rub.de}

\begin{abstract}
Conversational Artificial Intelligence (AI) systems have recently sky-rocketed in popularity and are now used in many applications, from car assistants to customer support. The development of conversational AI systems is supported by a large variety of software platforms, all with similar goals, but different focus points and functionalities. A systematic foundation for classifying conversational AI platforms is currently lacking. We propose a framework for assessing the maturity level of conversational AI development platforms. Our framework is based on a systematic literature review, in which we extracted common and distinguishing features of various open-source and commercial (or in-house) platforms. Inspired by language reference frameworks, we identify different maturity levels that a conversational AI development platform may exhibit in understanding and responding to user inputs. Our framework can guide organizations in selecting a conversational AI development platform according to their needs, as well as helping researchers and platform developers improving the maturity of their platforms.
\end{abstract}

%
%
%
%
%


\maketitle

\section{Introduction}

Conversational AI has recently surged in popularity and interest.
A conversational AI system is an interface that can communicate and interact with users by relying on the automated processing of questions and formulating answers. 
In 2016, Facebook announced a new platform to develop chatbots on their messaging application \cite{Menon2017CouncilAI}, which simplified the creation of AI chatbots by providing relevant toolkits \cite{MessengerPlatform}. After that, many other companies have implemented chatbots for both text and speech. Three of the most popular conversational AI systems today are Microsoft Cortana, Google Assistant, and Apple's Siri \cite{Tsai2018DataSpiceworks}.
The advances in deep learning \cite{yan2018chitty} and the advent of powerful language models, most recently GPT-3 \cite{DBLP:conf/nips/BrownMRSKDNSSAA20}, pave the way for a new generation of conversational AI systems enabling conversations with human-like qualities.


To support organizations in adopting conversational AI systems, a multitude of development platforms are available.
By offering numerous concepts, such as natural language understanding (NLU), webhooks, and contexts, these platforms enable the user to engineer systems that can provide a rich, ideally almost human-like conversation experience.
However, due to the large variety of available platforms, the relevance and need of each individual concept and its impact on the conversation experience is unclear. As a result, the use of such platforms may be overly complicated. 
To support organizations in selecting a suitable platform, and platform developers in increasing the maturity of their platforms, we need to improve our empirical understanding of the state-of-the art of the domain.
Specifically, we need to understand what platforms exist, what concepts they offer, what their concepts' characteristics are, in what combinations the concepts are used, and, in sum, what level of conversation they enable. Evaluating the conversational maturity that the different platforms offer might help in changing the perception that these systems are simply task-oriented tools, and that they can hold truly social conversations. Additionally, it may help in understanding how the more functional terms of these platforms relate to the conversational ability \cite{clark.ea:2019:challenges}.

In this paper, we provide a maturity assessment framework for conversational AI development platforms.
We provide a comprehensive overview of the features available in today's platforms, and
analyze these features to see how they relate to the quality and ability of conversational AI systems produced using them.
Finally, inspired by human language development frameworks, we propose a layered framework with multiple levels of conversational maturity.
With this contribution, we aim to improve our empirical understanding of current development platforms for creating conversational AI systems, their concepts, and the level of conversation that bots created with these systems can achieve.
As a benchmark for assessing existing and new platforms, our framework can support and guide practitioners who engineer such platforms.
Moreover, it can help researchers understand the concepts that exist, identify gaps between practice and research, and scope future research. In the long term, this could help in creating better conversational AI systems.

We address the following research questions:

\noindent\textbf{RQ1:} \emph{What platforms exist for developing conversational AI systems?}

\noindent\textbf{RQ2:} \emph{What are the features of these platforms?} 

These first two questions are aimed towards analyzing existing conversational AI development platforms and extracting information regarding their usage and features. A specific focus is on the ability of the platforms to model conversation dialogs.

To this end, we performed a literature study, in which we collected papers presenting different platforms. We then analyzed the documentation of the platforms to identify their distinguishing characteristics and concepts (features). To provide an intuitive, hierarchical overview on the multitude of available features, we grouped them into a feature model \cite{kang1990feature,czarnecki2000}, a common notation for modeling the variability of portfolios of software systems \cite{nevsic2019principles} in a domain. Feature models have also become popular in empirical studies for modeling the design space of technologies, such as model transformations\,\cite{transformationSurvey} or language-engineering workbenches\,\cite{erdweg2013languageworkbenches}.

\noindent\textbf{RQ3:} \textit{What are the levels of conversational maturity supported by the identified platforms?}

\looseness=-1
We created a framework that can be used to evaluate the conversational maturity (intuitively, how ``smart'' an agent is in understanding questions and formulating responses) offered by the platforms.
To this end, we considered existing frameworks that evaluate the language proficiency of humans, and previous discussions on how to evaluate different conversational AI development platforms. 
We then devised a framework based on the features identified in the first research questions and their effect on the human-like performance of a conversational AI development platform. 

\looseness=-1
There are not many studies on the conversational maturity that different conversational AI development platforms offer. The most closely related work is the study by Venkatesh et al. \cite{Venkatesh}, who describe how to evaluate the performance of conversational agents in terms of certain metrics. In contrast, our work focuses on recently available platforms and on how their features impact the conversational maturity of systems created upon these platforms (cf. Section~\ref{sec:relwork} for a discussion of related work).

\section{Background and Related Work}
\label{sec:relwork}

With the recent developments in many of the sub fields of conversational AI, including machine learning, dialog management and NLU, many different conversational AI systems have emerged \cite{Michiels2017ModellingExperience}. 

In industry, this technology has been incorporated into search engines, mobile devices, and personal computers. In search engines such as Google and Bing, conversational AI is used to create the feeling of having a conversation with the search engine, enhancing the experience. In mobile devices and personal computers, one use of conversational AI is to create virtual assistants. Some of the biggest virtual assistants on the market today are Apple’s Siri, Google Assistant, Amazon Alexa and Microsoft Cortana \cite{CunoKlopfenstein2017TheParadigms}. These assistants also have the capability of acting as chatbots where they keep a turn-based dialog (a dialog where the user and the bot take turns in asking and responding to queries) with the user. There also exist conversational interfaces that only focus on this type-dialog-based conversation such as XiaoIce \cite{Gao2018NeuralChatbots} and Replika \cite{Gao2018NeuralChatbots}. These dialogs use what is known within conversational AI as \emph{intents} and \emph{entities} to understand the user's goal behind the query. In other words, an \emph{intent} is what the user wants to achieve with the query, and an \emph{entity} is the key information for answering the intent.

Recently a number of different platforms have been made available to simplify the creation and integration of conversational interfaces for developers. The most popular ones are: Google's DialogFlow (formerly api.ai)\footnote{https://dialogflow.com/},
    IBM's cloud-based bot service Watson Conversation\footnote{https://www.ibm.com/cloud/watson-assistant/},
		Amazon Lex\footnote{https://aws.amazon.com/lex/}
    and the Microsoft Bot Framework\footnote{https://dev.botframework.com/}.
These platforms come equipped with several different technologies used for NLU, dialog management, response generation and other aspects \cite{Gao2018NeuralChatbots,Canonico2018ATools}.

Since conversational AI is a new field, systematic approaches to overview and categorize it are still in their infancy.
Patil et al. \cite{Patil2017} makes a general comparison of features and functionalities between some of the commercial platforms, giving an overview of what platform one might choose for developing a conversational AI system. There have also been more specific studies conducted which compare the NLU and conversational abilities of these types of platforms. Canonico and De Russis \cite{Massimo2018ATools} compare the NLU performance of these platforms have in terms of usability, pre-built intents (a number intents already existing in the NLU tool) context etc. McTear\cite{Mctear2018CONVERSATIONALDIRECTIONS} describe the two main conversation models "one-shot queries" and "slot-filling dialogues". He compares different platforms' ability to handle follow up questions in one-shot query scenarios and their mechanisms for slot-filling (a type of conversation where the bot asks specific questions to fill certain slots to fulfil a user intent). McTear also presents a number of problems that developers may face when creating conversational interfaces with these platforms. One of the main issues is that it might be difficult to know what functionalities a specific platform offers. There is also a difficulty in interpreting what functionalities might be common between platforms since there is no standard terminology.

Venkatesh et al. \cite{Venkatesh} describe a number of metrics that can be used to evaluate the overall performance of a conversational agent based on the annual competition Alexa Prize \cite{Ram2017} made for furthering conversational AI. They propose metrics such as conversational user experience, engagement, and conversational depth to measure the conversational abilities of entire conversational AI systems or chatbots \cite{Venkatesh}. Shawar and Atwell \cite{Shawar2007} describe metrics to specifically evaluate chatbot systems, a type of conversational AI interface. They argue that metrics for evaluating the abilities of these systems should be done based on the application and its domain and not solely on a standard. 

One of the main issues with creating the metrics described above is the understanding of what a good conversation is. Clark et al. \cite{clark.ea:2019:challenges} discuss that people generally describe conversations with conversational interfaces in terms of their performance and perceive them more as a device to be controlled. Indicating that people have a previous notion of how these systems will behave coming from a perception that infrastructure to support proper human-to-human dialogs do not exist.

The maturity assessment framework presented in this paper takes inspiration from three language proficiency frameworks: \textit{Common European Framework of Reference} (CEFR,~\cite{CouncilofEuropeCommonAssessment}), \textit{American Council on the Teaching of Foreign Languages} (ACTFL,~\cite{Swender2012ACTFL2012}), and the \textit{Interagency Language Roundtable} (ILR,~\cite{2019IteragencyRoundtable}).
The goal of these frameworks is to assess the language competency of an individual for a particular language.
All of these frameworks have a similar structure, distinguishing different, successive \textit{levels} (e.g., in case of CEFR, a six-item scale A1--C2), language-relevant \textit{skills} (e.g., for CEFR, \textit{reading}, \textit{listening}, \textit{speaking}, and \textit{writing}), and a number of hints for assigning an individual to a level. 
While the contents of the framework differ, they all share this same basic structure, which we also found useful for inspiring the design of our framework.
A number of papers have scientifically investigated these frameworks, studying their validity and the possibility to use them in an automated way \cite{Lange1987,Tschierner2012,Huang}. 

\section{Methodology}

\subsection{Identification of Conversational AI development platforms}
In the first part of our study (RQ1), we aimed to explore the variety of existing conversational AI development platforms. To this end, we used several methods as follows.

\subsubsection{Literature Review}
We used a systematic literature review to identify papers on conversational AI systems. We focused on papers that present and evaluate platforms used to develop such systems. 
Specifically, among the different methods that exist for conducting literature reviews, we used \textit{snowballing}. We followed Wohlin \cite{Wohlin2014GuidelinesEngineering}, who describes two types of snowballing and provides guidelines for performing them: forward and backwards snowballing. 
We performed backwards snowballing, to find papers describing current conversational AI development platforms.
Backwards snowballing involves selecting a number of papers to be used as a start set to find more relevant papers in the same field by tracing the reference lists of the papers. The start set should include a number of different papers from different areas of the field, different authors, and different points in time. The idea is to cover the considered field or topic to the largest possible extent. The reference lists of the papers in the start set are then evaluated based on certain inclusion and exclusion criteria (explained shortly). From the start set, additional papers can be found, which we also screened. Each set of reviewed papers is one iteration of the snowballing procedure, once no more papers can be found the process is over. \cite{Wohlin2014GuidelinesEngineering}

We collected the start set for snowballing through database searches, using search strings such as ``Conversational AI,'' ``Conversational AI development platforms,'' and ``Chatbot platforms''. We provide the full list as supplementary material (made available in our online appendix \cite{onlineappendix}). The first 50 results for each search string were examined based the criteria listed below. To determine whether to include or exclude a certain paper we used the following procedure: Read the title and abstract and skim the whole paper to determine if any relevant platforms can be found. A paper could be excluded at any stage of the process based on its relevance to the study. The databases used in this search were Google Scholar, IEEExplore, arXiv, SpringerLink, and a university library database.

Our inclusion criteria were:
\begin{itemize}
\compactlist
    \item Papers published after 2000, after which most recent platforms have been developed, were candidates for inclusion.
    \item Papers examining and presenting different conversational AI/bot platforms were included.
    \item Papers that only examine characteristics of conversational AI and do not mention any platforms were excluded.
\end{itemize}

The platforms that were found through the literature review were then examined in order to determine if enough information about them was available to fairly assess what features the platforms provided.

To this end, our exclusion criteria were:
\begin{itemize}
\compactlist
    \item Platforms that are no longer available or heavily outdated (no update since two years at time of search) were excluded.
    \item Systems that were just simple chatbots (i.e., responding only to simplest queries like \textit{"tell me the time"}) were excluded.
    \item Platforms that did not have enough documentation available publicly were excluded.
    \item Platforms that did not have a strong enough user base, either by individuals or companies or both, were excluded.
\end{itemize}

In addition to conducting snowballing, we consulted with an employee from an industrial partner---a company
with years of experience in conversational AI---to find platforms that we might have missed. This made us aware of several additional platforms (detailed below).
We conducted our analysis  in the summer months of 2019.

\subsubsection{Database Searches}
To find platforms outside the more formal channels of published literature and company expertise, we also sought via the Google search engine. This required specific care and source criticism, since the information available may be outdated or even false. We conducted the searches using search terms that try to find platforms similar to those found through previous methods, for instance, ``DialogFlow competitors.''

\subsection{Documentation Analysis}
The main process for collecting information about the different conversational AI development platforms was document analysis. Document analysis involves going trough any documentation available for a specific entity, such as a software platform. It allows for the collection of data that later can be evaluated and grouped based on certain criteria. Document analysis is often quite efficient and cost-effective since no new data needs to be acquired; instead, already existing data is evaluated. However, there is a risk that the documentation may be incomplete~\cite{Bowen2009DocumentMethod}.

We analyzed the documentation available for all considered platforms to identify their common and distinguishing features, thus addressing RQ2.  Whilst the entire platforms were analyzed to be able to give an overview of the entire system structure, we put special emphasis on their \textit{conversation-defining features}. The conversation-defining features build up the dialog management portion of the platforms, which define what the bot can understand and how it should respond. 
This process also helped us mapping similar features whose names vary between different platforms.

\subsection{Feature Model}
To represent the identified features, we developed a feature model~\cite{kang1990feature}. Feature models visualize the features of a platform by displaying them in a hierarchy, thus providing a good overview of top-level and more fine-grained features. Features can be mandatory or optional. 
In our survey, we refer to common features of the considered platforms as mandatory, and to distinguishing ones as optional. The model also includes constraints between the features, such as \textit{dependencies}, in which a feature needs another feature for its implementation.
There are a few other models that can be used for similar purposes, such as class diagrams. However, we used feature models since they provide a compact, hierarchical overview, which is good for managing complexity in large systems \cite{Lee2002ConceptsEngineering}, and since feature models have been used in earlier empirical studies \cite{transformationSurvey,erdweg2013languageworkbenches} on systematizing the features of systems in a particular domain.


\subsection{Designing the Conversational AI Maturity Framework}

\subsubsection{Identification of Language Maturity Frameworks}
As a prerequisite for creating a maturity framework for conversational AI development platforms, we explored if any similar attempts had been made before.
We performed a literature review to identify any existing frameworks, either directly related to conversational AI classification or to evaluate the conversational maturity of a human.
We searched using Google Scholar, IEEE, arXiv, Springer and our university's library.
The following search phrases were used when looking for these frameworks: ``Common language framework,'' ``Human language framework,'' and ``Language framework. More can be found in the online appendix \cite{onlineappendix}. From these searches, the top 50 results were considered to determine their relevance for this study. We used the following exclusion and inclusion criteria:

\begin{itemize}
\compactlist
    \item Papers discussing different aspects of what makes good conversational AI were included.
    \item Papers with frameworks used to evaluate maturity of either human or bot conversation maturity were included.
    \item Papers that have metrics for evaluating conversational AI systems were included.
\end{itemize}

To determine whether a paper matched the criteria above the following procedure was followed: read the title of the papers; read the abstract of the papers; read discussion to determine if any frameworks are presented or characteristics of good conversational AI are mentioned. As mentioned above a paper could be excluded at any point of the process, if the title was out of scope the paper is directly excluded.

\subsubsection{Designing the Framework}
\label{sec:frameworkdesign}
\looseness=-1
Our goal was to create a framework that describes a collection of incremental levels of conversational maturity, inspired by the language proficiency frameworks CEFR, ACTFL, and ILR.
We used the same structure as these existing frameworks (distinguishing various incremental \textit{levels} and orthogonal \textit{skills}), but filled the structure with entirely new contents, tailored to our understanding of conversational maturity.

Since a definition of \textit{conversational maturity} was not available in the literature, we devised an own definition:
\textit{The ability of a conversational AI system to participate in a human-like conversation}.
To identify levels, we used the features found through the documentation analysis.
We decided for each feature if it contributes to conversational maturity according to this definition, and clustered those features that do into distinct, progressive levels.

\section{Characterising Conversational AI development platforms}
\label{sec:features}
We present the results from our literature review, the documentation analysis performed on the identified conversational AI development platforms, and the obtained feature model with common and distinguishing features of these platforms.

\subsection{Result of Literature Review}

The database searches resulted in 10 sources which were used as the start set for snowballing. The references of each of these papers were then screened in order to find any other papers relevant for the purpose of finding conversational AI development platforms. The snowballing was ended after three iterations of this procedure. \\

In the first iteration, based on the start set, 13 additional papers were added. The list of potential candidates were narrowed down by using the following procedure: Read title of papers; check where the paper is referenced in the text; read abstract of papers; look at full text to determine if it contains any new conversational platforms. The place of reference was checked in order to determine if it was used in conjunction with text that describe conversational platforms. All papers were matched against the same criteria that was used to put together the start set, see Section 3. The second iteration of the snowballing procedure were done on the 13 newly found papers. From these papers another 3 were identified that describe conversational AI development platforms. The third and last iteration identified no additional relevant papers.

Using snowballing, we identified a total of 56 different potential conversational AI development platforms.
From these 56 platforms, we removed a number of duplicates arising from the same system appearing under different names:
DialogFlow was renamed to API.ai, and IBM voice server and AT\&T watson were the predecessors to IBM Watson Conversation. 
We excluded the conversational interfaces Cortana, Google assistant, and Amazon Alexa, as they are not actual development platforms. Cortana is developed by Microsoft who makes its technology available through Microsoft Bot Framework. Google assistant is very much related to DialogFlow and Amazon Alexa with Amazon Lex.
The remaining platforms were matched against the inclusion and exclusion criteria mentioned in Section 3. These criteria were used to narrow down the set to a total of 10 platforms:
DialogFlow, Microsoft Bot Framework, Houndify, RASA, Amazon Lex, IBM Watson Conversation, VoiceXML, Recast.ai, Kore.ai, and AIML. 



Consulting with an employee of our partner company resulted in the addition of three new platforms that had not yet been examined as well as the confirmation that the platforms found in the literature review match many of the platforms that had been found by the company. The three new platforms that were found are: Teneo, Boost.ai, and TDM. Teneo and Boost.ai were not included in further investigations as they lacked sufficient documentation of their contained features.


Lastly, we used the Google search engine to identify potentially missed conversational AI development platforms. 
Our search brought forwad three new platforms that had emerged quite recently: Meya, Chatbot and Botpress. Chatbot and Botpress lacked available documentation supporting a fair assessment of the available functionality. For this reason both of these platforms were excluded. Meya was included since its documentation was extensive enough to form a full image of its features.

We finally obtained a list of twelve platforms to further analyze. 
Table~\ref{tbl:platforms} shows these platforms, together with their core characteristics: open- vs. closed-source, (semi-)commercial vs. free, and web-based vs. command-line vs. implementation-dependent.
A platform is semi-commercial if it has both free and paid variants, where the free variant typically has an upper cap (for example, 10,000 messages per month in the Microsoft Bot Framework).

\begin{table}[t]
\caption{Conversational AI development platforms}
\label{tbl:platforms}
\begin{tabular}{m{0.11\textwidth} | m{0.05\textwidth} | m{0.095\textwidth} | m{0.13\textwidth}} \toprule
\label{framework}
\textbf{Platform} & \textbf{Source} & \textbf{Availability} & \textbf{Modality}\\ \hline
DialogFlow & closed & commercial & web-based\\ \hline
Meya.ai & closed & commercial & web-based\\ \hline
Microsoft Bot Framework & closed & semi-comm. & web-based\\ \hline
Houndify & closed & semi-comm. & web-based\\ \hline
Amazon Lex & closed & commercial & web-based\\ \hline
RASA & open & free & command-line\\ \hline
IBM Watson Conversation & closed & semi-comm. & web-based\\ \hline
VoiceXML & open & free & impl.-dependent\\ \hline
Recast.ai & closed & semi-comm. & web-based\\ \hline
Kore.ai & closed & semi-comm. & web-based\\ \hline
AIML & closed & free & impl.-dependent\\ \hline
TDM & closed & commercial & command-line\\ \bottomrule
\end{tabular}
{\footnotesize URLs: https://cloud.google.com/dialogflow
https://meya.ai/
https://botframework.com/
https://houndify.com/
https://aws.amazon.com/lex/
https://rasa.com/
https://ibm.com/cloud/watson-assistant/
https://w3.org/Voice/
https://recast.ai
http://aiml.foundation/
http://talkamatic.se/}
\end{table}

\subsection{Results of the Documentation Analysis}
We thoroughly analysed the documentation of platform to pinpoint its included functionality, resulting in a list of features.
These features were then reviewed to identify common and distinguishing features between the different platforms. If the same feature was found in multiple platforms under different names, we continued with the name used more often in the platforms and existing literature. The consolidated features were added to a feature matrix (made available in our online appendix \cite{onlineappendix}).
The end result was a list of 54 different features that we grouped and organized in a hierarchy to obtain a feature model, discussed in what follows.



\subsection{Feature Model}
\begin{figure}[t]
    \centering
    \includegraphics[scale=0.88, trim = 0cm 26.35cm 10cm 0cm, clip]{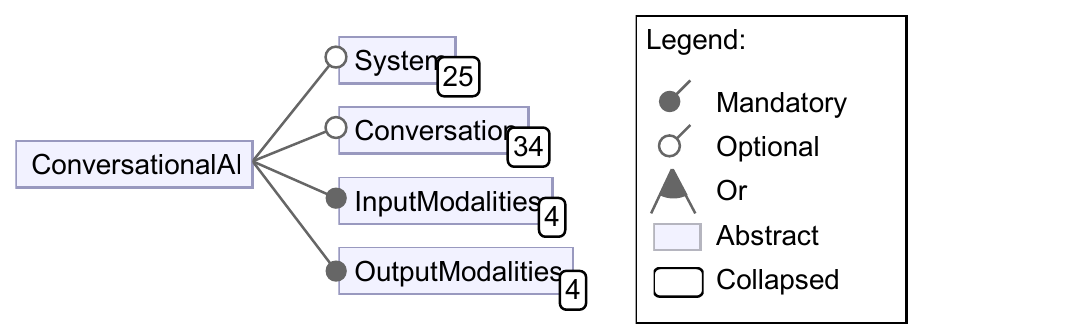}
		\vspace{-15pt}
    \caption{Top level view of the feature model}
    \label{fig:toplevel}
\end{figure}

Figure~\ref{fig:toplevel} shows a high-level overview of our feature model,
highlighting its four top-level feature groups:
\textsl{System}, \textsl{Conversation}, \textsl{Input modalities}, \textsl{Output modalities}.
These top-level feature groups and their contained features will be discussed in this section.
The full list of all features with their descriptions is made available in our online appendix \cite{onlineappendix}.

We use the standard syntax of feature models.
Specifically, as shown in the legend, features can be marked as mandatory, meaning that they exist in all or most of the analyzed platforms, or optional, meaning that they only exist in some platforms.
Numbers attached to a node indicate that the node, in fact, represents a collapsed sub-tree, with the specified number of total nodes in the sub-tree.
Abstract features are used for grouping purposes.
"Or" features are used to specify features groups, where each considered platform had at least one of several features of the group.
In what follows, we describe the most crucial features, including those deemed as particularly relevant for assessing conversational maturity.

\begin{figure}[t]
    \centering
    \includegraphics[scale=0.7, trim = 0cm 23.22cm 8cm 0cm, clip]{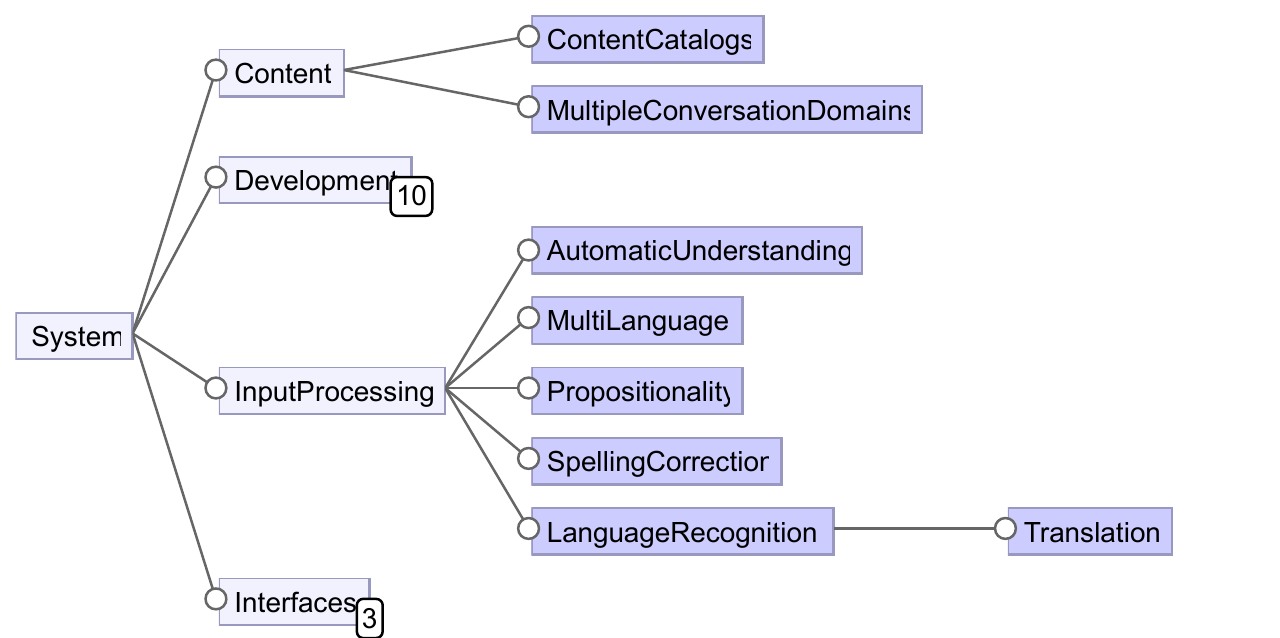}
		\vspace{-15pt}
    \caption{Main system features}
    \label{fig:systemfeatures}
\end{figure}

\subsubsection{System Features}
From multiple \textit{System} features as shown in Figure~\ref{fig:systemfeatures}, \textit{Content} refers to the different conversation contents and how the platform handles them. A conversation content can be, for example, "phoning a friend" or "weather information". Figure~\ref{fig:systemfeatures} shows two crucial sub-features of content:
\textit{ContentCatalogs} refers to platforms with in-built content catalogues to simplify the development of the conversational AI bot. These catalogs
contain entities and intents that are common in the domain.
\textit{MultipleConversationDomains} is used to distinguish platforms that support the handling of domains that are completely independent from one another, thus making it possible to have 2 different content domains in the same conversation.
This is in contrast to most platforms, which only support one particular domains with no explicit separation from other domains.
\textit{MultiLanguage} is the conversational AI feature that regards to the number of supported languages within the platform.

\textsl{Development} features concern the development process of systems using the platform, which can be supported by features such as error feedback, debugging, and versioning tools.
\textit{Input processing} refers to features to processing of the user input, such as \textit{SpellingCorrection}.
Different \textit{interfaces} being supported may include a custom frontend, integration with social media, and other websites.


\subsubsection{Conversation Features}
\begin{figure}[t]
    \centering
    \includegraphics[scale=0.9, trim = 0cm 23.55cm 14cm 0cm, clip]{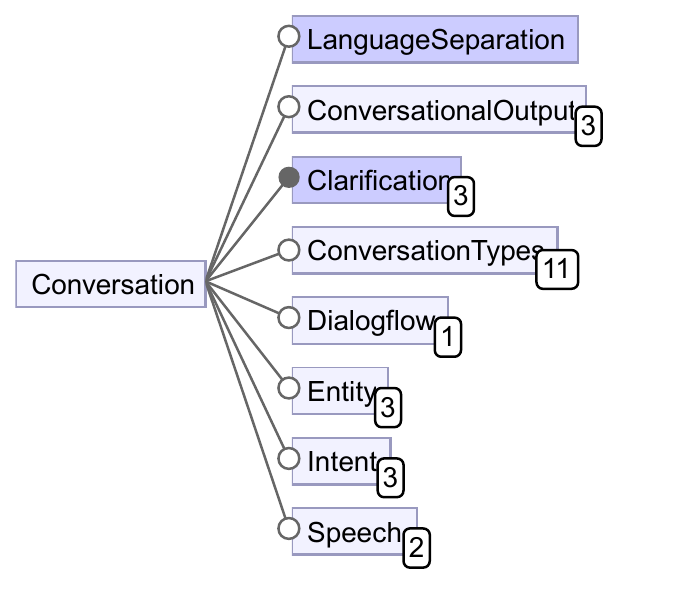}
		\vspace{-15pt}
    \caption{Conversational features}
    \label{fig:conversationalfeatures}
\end{figure}

One of the main reasons why there are so many different conversational AI development platforms is that most handle conversations differently from one another. These differences can be anything from the content of the conversation to the dialog management of the platform. The platforms consider conversations differently depending on what the intent of use is and the area of use is. Many of the different platforms in the market are focused on one specific field of expertise and are customized to fit the needs and standards of this field. An overview of these features and feature categories can be seen in Fig.~\ref{fig:conversationalfeatures}. These features are all regarding the conversation between the agent and a user, everything from processing to supported conversation types. Language-specific features are also in the \textit{Conversation} category, since the language is a part of the conversation.

\textit{LanguageSeparation} is the ability of the system to separate language-specific and non-language-specific information from a sentence. This allows the system to identify what parts of the sentence is crucial for the information to be received and what parts are not. This feature will simplify the translation of a conversation and the multilingual maintenance of the system. 

\begin{figure}[t]
    \centering
    \includegraphics[scale=0.88, trim = 0cm 27.5cm 14cm 0cm, clip]{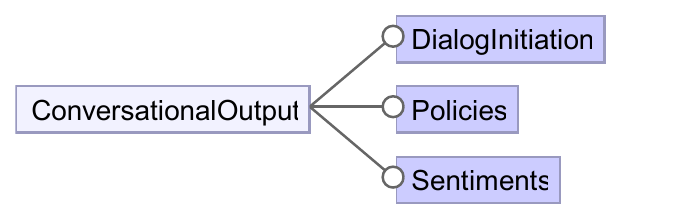}
		\vspace{-10pt}
    \caption{Conversational output features}
    \label{fig:conversationaloutputfeatures}
\end{figure}

\textit{Conversational output} features, depicted in Fig.~\ref{fig:conversationaloutputfeatures}, affect how the conversational output is processed. \textit{DialogInitiation} is one way to do so, it allows the developer to instigate a conversation. Different companies have different \textit{Policies} and rules to adhere so some platform allows for such policies and rules to be implemented within the system it self. \textit{Sentiments} allows for the conversational AI to display emotions in their response.

\begin{figure}[t]
    \centering
    \includegraphics[scale=0.84, trim = 0cm 27.45cm 14cm 0cm, clip]{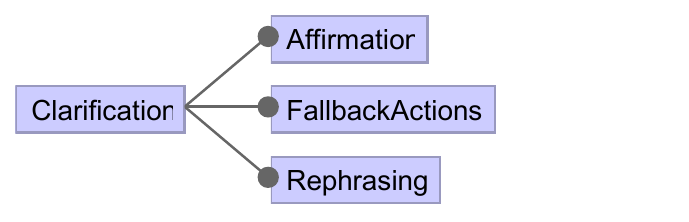}
		\vspace{-10pt}
    \caption{Clarification features}
    \label{fig:clarificationfeatures}
\end{figure}

\textit{Clarification} is something that all the analyzed platforms support in some way, since not misunderstanding the user is crucial for supporting the robustness of the system. As summarized in Fig.~\ref{fig:clarificationfeatures}, this is done by using \textit{Affirmation}, \textit{Rephrasing} or \textit{FallbackActions} to confirm the users intent. These features affect how the system reacts when a user input is not understood or if a user input can be assigned to two different intents. These features also allow to give the user a second chance to change their mind or query. 

\begin{figure}[t]
    \centering
    \includegraphics[scale=0.62, trim = 0cm 24.03cm 5cm 0cm, clip]{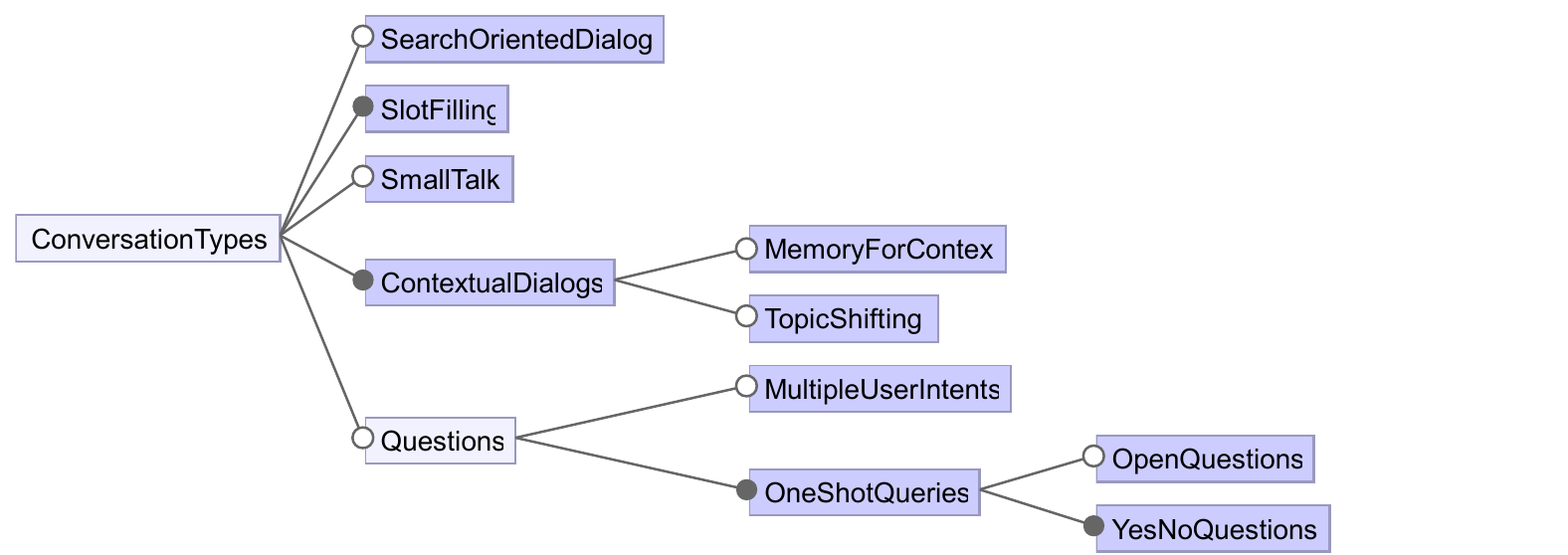}
		\vspace{-15pt}
    \caption{Conversation types}
    \label{fig:conversationtypes}
\end{figure}

\textit{ConversationTypes} features cover the different types of conversation and questions that a platform supports. These features and feature subsets can be seen in Figure~\ref{fig:conversationtypes}.

\textit{SearchOrientedDialog} refers to a dialog that searches through a database to find matching entities and respond to a user intent.

\textit{SlotFilling} is a dialog where to conversational AI ask for additional information to fill certain criteria to match the correct intent to an entity. An example of this could be:
    \begin{itemize}
    \compactlist
        \item [User: ] What is the weather like today?
        \item [Bot: ] Which city would you like to search the weather for?
        \item [User: ] New York.
        \item [Bot: ] The weather in New York is cloudy.
 
    \end{itemize}

\textit{SmallTalk} is a conversation type that refers to conversations without any specific end goal. These types of conversations can be anything from asking how you feel to telling jokes.    
    
\textit{ContextualDialogs} are an important component  of conversational AI systems, supported by every analyzed platform. We found two different ways to support contextual dialogs: one or multiple contexts per conversation. To support contextual dialogs a conversational AI development platform must have the feature \textit{MemoryForContext}. MemoryForContext is specially allocated memory that the AI uses in order to remember previous information.
Multiple context are referred to as \textit{TopicShifting} and can enable conversations like, for example:
    \begin{itemize}
    \compactlist
        \item [User: ] Send a text message to Peter.
        \item [Bot: ] What would you like to text?
        \item [User: ] I want to book a flight to Japan for tomorrow.
        \item [Bot: ] What time would you like to book the flight at tomorrow?
        \item [User: ] At 4am.
        \item [Bot: ] Where in Japan would you like to fly?
        \item [User: ] I would like to text: "The temperature in New York is 20$^{\circ}$C."
        \item [Bot: ] Ok, your message has been sent.
        \item [User: ] I would like to fly to Tokyo.
        \item [Bot: ] Ok, flight booked to Tokyo tomorrow at 4am.
    \end{itemize}

The last conversation type is \textit{Questions}, these can vary from simple \textit{OneShotQueries} to complex \textit{MultipleUserIntents}. There are two types of OneShotQueries: \textit{YesNoQuestions} and \textit{OpenQuestions}. Both these types only require one response from the conversational AI to fully answer the query. MultipleUserIntents are queries with multiple intents within them, an example can be: "What is the time and weather like in New York?"
    
\begin{figure}[t]
    \centering
    \includegraphics[scale=0.8, trim = 0cm 28.2cm 13cm 0cm, clip]{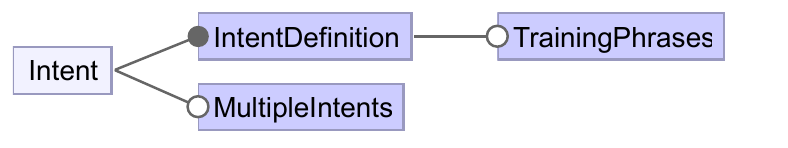}
		\vspace{-10pt}
    \caption{Intent features}
		\vspace{-10pt}
    \label{fig:intentfeatures}
\end{figure}

Features related to \textit{Intent} are concerned with intent manipulation and intent restrictions. Intents are used to define the users goal with the query, for example:
    \begin{itemize}
    \compactlist
        \item [User: ] What is the time in New York?
        \item [Bot: ] The time in New York is 1pm.
    \end{itemize}
In this case the intent of the user is: finding out the time. These features can be seen in Figure~\ref{fig:intentfeatures}. 

\begin{figure}[t]
    \centering
    \includegraphics[scale=0.80, trim = 0cm 28.15cm 14cm 0cm, clip]{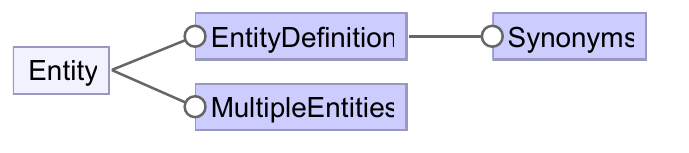}
		\vspace{-10pt}
    \caption{Entity features}
		\vspace{-10pt}
    \label{fig:entityfeatures}
\end{figure}

\textit{Entity} features, as shown in Figure~\ref{fig:entityfeatures}, are concerned with entity manipulation and entity restrictions. Entities are descriptive actions the conversational AI can perform after identifying the users intent. An example is:
\begin{itemize}
    \compactlist
        \item [User: ] Can you call mom?
        \item [Bot: ] Calling mom.
\end{itemize}
In this case the intent would be to make a call and the entity would be mom.

\begin{figure}[t]
    \centering
    \includegraphics[scale=0.8, trim = 0cm 28.3cm 14cm 0cm, clip]{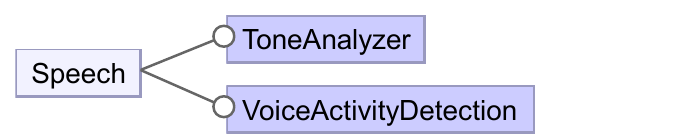}
		\vspace{-10pt}
    \caption{Speech features}
		\vspace{-10pt}
    \label{fig:speechfeatures}
\end{figure}

Several platforms support features specific to \textit{Speech}, one such feature is \textit{VoiceActivityDetection} which allows the system to detect changes in audio level to determine whether or not the user is currently speaking. Another feature is \textit{ToneAnalyzer} which allows the conversational AI to identify emotions within the speech pattern. These features, depicted in Fig.~\ref{fig:speechfeatures}, are required by those platforms that support speech as an input modality. 

\subsubsection{Input Modalities}
\begin{figure}[t]
    \centering
    \includegraphics[scale=0.77, trim = 0cm 26.9cm 15.9cm 0cm, clip]{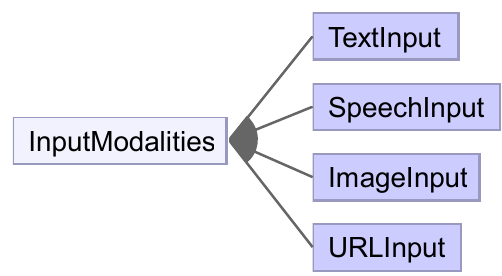}
    \includegraphics[scale=0.77, trim = 0cm 26.9cm 15.5cm 0cm, clip]{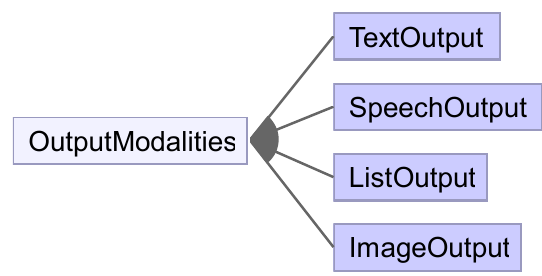}
		\vspace{-4pt}
    \caption{Input and output modalities}
    \label{fig:input}
\end{figure}

For a user to communicate with a conversational system the platform must support reading and analyzing one or several input types. The different input modalities supported by the considered platforms are shown in Figure~\ref{fig:input}: input of text, speech, images, and URLs.  The larger the number of input modalities a platform supports, the larger the potential areas of use and accessibility of the created systems. 

\subsubsection{Output Modalities}
For a two-sided conversation, the conversational AI system must be able to answer to respond to user queries by using one or several output modalities, the output modalities can also be seen in Fig.~\ref{fig:input}. These different types of output, like input modalities, allow the system to be used in different types of environments. Some environments only allow for one type of output to not disrupt the user's focus. 
For example, in a moving car,  the optimal type of output would be speech.
To have several types of output to choose from will both enhance the usability and the accessibility of the developed system.

\section{Maturity Assessment Framework}
\label{sec:framework}

To support the evaluation of the conversational maturity of a conversational AI development platform, we created a maturity framework. The framework is inspired by those for human language development such as CEFR, ILR and ACTFL, which help to assess the conversational maturity of a language learner. This maturity framework can be used in the same fashion, to evaluate the  conversational maturity level of a conversational AI development platform and how it compares to other platforms.
We obtained the framework by considering all features of the state-of-the-art systems surveyed in Section \ref{sec:features}, using the methodology outlined in Sect.~\ref{sec:frameworkdesign}.

\subsection{Overview}
Our framework, as summarized in Table \ref{tbl:framework}, is divided into four main levels, where each level is further divided into two sub-levels for further differentiation (inspired by the organization of the CEFR into three main levels with two sub-levels each).
Orthogonally to the levels, the framework distinguishes two main skills, targeted to the capabilities of conversational AI systems: \textit{understanding} and \textit{response}. Understanding refers to the level of comprehension the conversational AI system has and the type of natural language processing it can perform. Response refers to the system's response patterns and abilities to interact with the user.

Table \ref{framework} further specifies the features corresponding to each level, pinpointing how certain functionality enables the conversational maturity of a conversational AI development platform.
All relevant features have been introduced in Section \ref{sec:features}. 
For a conversational AI development platform to be assigned as a specific level it has to accommodate all the preceding levels of conversational maturity.
For example: to be assessed as level 2a, the platform must satisfy all criteria for levels 1a and 1b, in addition to those of 2a.

\subsection{Maturity levels}

A conversational AI development platform's maturity level depends on its conversational abilities.
We propose the following levels:

\begin{description}
\compactlist
    \item[Level 1] Indicates very limited conversational abilities and very simple                 comprehension.
        \begin{description}
        \item[Level 1a] Capabilities within the explicitly defined domain and response                 restricted to isolated words.
        \item[Level 1b] Capabilities within related domains and response restricted to                 sentence fragments.
        \end{description}
    \item[Level 2] Indicates abilities to hold a short conversation with context and             comprehension on a intermediate level.
        \begin{description}
        \item[Level 2a] Capabilities to understand longer queries and responding with full             sentences.
        \item[Level 2b] Capabilities to memorize context for a short conversation and can             respond with questions.
        \end{description}
    \item[Level 3] Indicates abilities to hold a long conversation with context and             comprehension on an advanced level.
        \begin{description}
        \item[Level 3a] Capabilities to comprehend spelling mistakes and small talk.
        \item[Level 3b] Capabilities to comprehend multiple intents in one query.
        \end{description}
    \item[Level 4] Indicates abilities to hold multiple conversations and comprehending         complex human features.
        \begin{description}
        \item[Level 4a] Capabilities to comprehend multiple input languages and responding             in using different languages.
        \item[Level 4b] Capabilities to comprehend feelings and sentiments and use it for             responding.
        \end{description}
\end{description}


Level 1 describes a platform's ability to understand and respond to simple one-shot queries and phrases, which requires support for simple \textit{intent} and \textit{entity} definition as well as basic \textit{dialog management} based on a definition of isolated words. It also needs to be able to provide some basic \textit{input} and \textit{output processing}. 

Level 2 is devoted to intermediate-level conversation.
A feature that significantly affects the conversational maturity of a conversational AI is \textit{context}. In conjunction with other features it plays a very crucial role in creating more fluid conversations between the AI and a user. 

Level 3 addresses advanced conversational ability.
A complex feature of conversational AI is the ability to process and understand multiple intents within one query,  corresponding to level 3b of the framework. This would not only require all the aforementioned features, but also needs features such as: \textit{multiple intents per entity} and \textit{nested intents}. Making it possible to handle many different user requests at once makes for a more human-like conversation.

Level 4 considers one of the most important and challenging factors of human conversation, the interpretation and understanding of emotions. A conversational AI system that can both read and respond with feelings creates a more natural and as such a more mature conversation, allowing a proper relationship with the user to be developed \cite{Leite2013,Mctear2016TheDevices}. In conversational platforms this functionality is supported through the features \textsl{sentiments} and \textit{tone analysis}. This together with features such as \textit{context}, \textit{multiple intents} and \textit{multi-domains} make for a platform that can support some of the most mature conversational AI systems. 

\begin{table}[t!]
\caption{Maturity levels of considered platforms (based on our analysis in 2019; possible later updates not yet included)}
\label{tbl:applied}
\begin{tabular}{lll}
\toprule
\multicolumn{1}{l}{\textbf{Platform}}                & \multicolumn{1}{l}{\textbf{Lv.}} & \multicolumn{1}{l}{\textbf{Features for next level}}              \\ \hline
\multicolumn{1}{l}{DialogFlow}              & \multicolumn{1}{l}{3b}                & \multicolumn{1}{l}{TopicShifting, Sentiments, Policies}        \\ \hline
\multicolumn{1}{l}{Meya.ai}                 & \multicolumn{1}{l}{1b}                & \multicolumn{1}{l}{OpenQuestions, MultiConv.Dom.} \\ \hline
\multicolumn{1}{l}{Microsoft Bot Fw.} & \multicolumn{1}{l}{3a}                & \multicolumn{1}{l}{MultipleUserIntents}                        \\ \hline
\multicolumn{1}{l}{Houndify}                & \multicolumn{1}{l}{1b}                & \multicolumn{1}{l}{SearchOrientedDialog}                       \\ \hline
\multicolumn{1}{l}{Amazon Lex}              & \multicolumn{1}{l}{3b}                & \multicolumn{1}{l}{MultiLanguage, Policies}                 \\ \hline
\multicolumn{1}{l}{RASA}                    & \multicolumn{1}{l}{2b}                & \multicolumn{1}{l}{SpellingCorrection}                         \\ \hline
\multicolumn{1}{l}{IBM Watson Conv.} & \multicolumn{1}{l}{3a}                & \multicolumn{1}{l}{LanguageSeparation}                         \\ \hline
\multicolumn{1}{l}{VoiceXML}                & \multicolumn{1}{l}{3a}                & \multicolumn{1}{l}{LanguageSeparation}                         \\ \hline
\multicolumn{1}{l}{Recast.ai}               & \multicolumn{1}{l}{1b}                & \multicolumn{1}{l}{SearchOrientedDialog}                       \\ \hline
\multicolumn{1}{l}{Kore.ai}                 & \multicolumn{1}{l}{1b}                & \multicolumn{1}{l}{SearchOrientedDialog}                       \\ \hline
\multicolumn{1}{l}{AIML}                    & \multicolumn{1}{l}{1b}                & \multicolumn{1}{l}{SearchOrientedDialog}                       \\ \hline
\multicolumn{1}{l}{TDM}                     & \multicolumn{1}{l}{2a}                & \multicolumn{1}{l}{MemoryForContext}                           \\ \bottomrule                                                                
\end{tabular}
\vspace{-10pt}
\end{table}

\begin{table*}
\caption{Assessment framework for conversational AI development platforms: maturity levels.}
\label{tbl:framework}
\begin{tabular}{m{0.07\textwidth} | m{0.35\textwidth} | m{0.27\textwidth} | m{0.2\textwidth}} \toprule
\label{framework}
\textbf{Level} & \textbf{Understanding} & \textbf{Response} & \textbf{Features}\\ \midrule
\textbf{Level 1 } & \multicolumn{3}{l}{Very limited conversational abilities and very simple comprehension}\\ \midrule
\textbf{Level 1a} &
\begin{itemize}[leftmargin=*]  
 \compactlist
	\item Can understand simple one-shot queries, like yes-or-no 			questions and questions for previously defined names. 
	\item Can understand simple phrases like greetings or information regarding 		the domain at hand.
	\item Can understand one intent per entity defined in the conversational AI.
\end{itemize} \nointerlineskip  & 
\begin{itemize}	[leftmargin=*]
\compactlist 
	\item Can respond with isolated words and numbers like “yes”, “no” and “50”.
	\item Can initiate a dialog with the user, instead of waiting for a user 			input.
\end{itemize} \nointerlineskip  & 
\begin{itemize}[leftmargin=*]
\compactlist 
	\item Intent
	\item Entity
	\item OneShotQueries
	\item YesNoQuestions
	\item InputModality
	\item OutputModality
	\item DialogInitiation 
\end{itemize} \nointerlineskip \\ \midrule
\textbf{Level 1b} &
\begin{itemize}[leftmargin=*]
\compactlist  
	\item Can understand queries that have been explicitly 	defined for the domain at hand.
	\item Can also understand simple queries for related domains.
\end{itemize} \nointerlineskip  & 
\begin{itemize}[leftmargin=*]  
\compactlist
	\item Can build a short phrase or sentence using a few connected words, to 			produce and response.
\end{itemize} \nointerlineskip  & 
\begin{itemize}[leftmargin=*]
\compactlist 
	\item OpenQuestions
	\item MultipleConver\-sationDomains
\end{itemize} \nointerlineskip  \\ \hline \\\\[-3\medskipamount]
\textbf{Level 2 } & \multicolumn{3}{l}{Abilities to hold a short conversation with
context and comprehension on a intermediate level}\\ \midrule
\textbf{Level 2a} &
\begin{itemize}[leftmargin=*] 
\compactlist
	\item Can understand longer queries with nested intents within the defined 			domain. 
	\item Can understand queries for related domains.
\end{itemize} \nointerlineskip  & 
\begin{itemize}[leftmargin=*] 
\compactlist 
	\item Can respond with full sentences containing more information.
	\item Can use affirmation to confirm the users intent.
	\item Can ask the user to repeat the query if it was not understood or 		misheard.
\end{itemize} \nointerlineskip  & 
\begin{itemize}[leftmargin=*]
\compactlist 
	\item Affirmation
	\item Rephrasing
	\item FallbackActions
	\item SearchOrientedDialog
	\item SlotFilling
\end{itemize} \nointerlineskip \\ \midrule
\textbf{Level 2b} &
\begin{itemize}[leftmargin=*]  
\compactlist
	\item Can understand the conversational context and keep that context 				memorized throughout a short conversation. 
	\item Can understand multiple intents for a single entity defined in the 			conversational AI.
	\item Can comprehend propositionality.
\end{itemize} \nointerlineskip  & 
\begin{itemize}[leftmargin=*]
\compactlist
	\item Will ask the user for additional information if there is missing 				information for the intent.
	\item Can respond with follow up questions to further continue the 					conversation.
\end{itemize} \nointerlineskip  & 
\begin{itemize}[leftmargin=*]
\compactlist 
	\item ContextualDialogs
	\item MemoryForContext
\end{itemize} \nointerlineskip \\ \hline \\\\[-3\medskipamount]
\textbf{Level 3 } & \multicolumn{3}{l}{Abilities to  hold a long conversation with
context and comprehension on an advanced level}\\ \midrule
\textbf{Level 3a} &
\begin{itemize}[leftmargin=*]  
\compactlist
	\item Can understand context in a sentence and keep that context memorized 			throughout an entire conversation. 
	\item Can use spelling correction and understand policies and censorship.
	\item Can comprehend small talk, like asking: “How are you?”
\end{itemize} \nointerlineskip  & 
\begin{itemize}[leftmargin=*]
\compactlist
	\item Can respond with sentences regarding small talk, like: “I’m doing very 		good!”.
\end{itemize} \nointerlineskip  & 
\begin{itemize}[leftmargin=*]
\compactlist 
	\item SpellingCorrection
	\item SmallTalk
\end{itemize} \nointerlineskip \\ \midrule
\textbf{Level 3b} &
\begin{itemize}[leftmargin=*] 
\compactlist
	\item Can understand multiple intents in one query.
	\item Can separate between language-specific and non-language-specific 				information.
\end{itemize} \nointerlineskip  & 
\begin{itemize}[leftmargin=*] 
\compactlist
	\item Can respond with multiple answers to cover all intents in the query.
\end{itemize} \nointerlineskip  & 
\begin{itemize}[leftmargin=*]
\compactlist 
	\item MultipleUserIntents
	\item LanguageSeparation
\end{itemize} \nointerlineskip \\ \hline \\\\[-3\medskipamount]
\textbf{Level 4 } & \multicolumn{3}{l}{Abilities to hold multiple conversations and
comprehend complex human language features}\\ \midrule
\textbf{Level 4a} &
\begin{itemize}[leftmargin=*] 
\compactlist 
	\item Can shift between different contexts within the same conversation.
	\item Can understand at least 2 input languages.
\end{itemize} \nointerlineskip  & 
\begin{itemize}[leftmargin=*]
\compactlist 
	\item Can have sentiments when answering queries to add a more human aspect. 		Can respond in different languages.
	\item Can translate information for the user.
\end{itemize} \nointerlineskip  & 
\begin{itemize}[leftmargin=*]
\compactlist 
	\item TopicShifting
	\item MultiLanguage
	\item Sentiments
	\item Policies
\end{itemize} \nointerlineskip \\ \midrule
\textbf{Level 4b} &
\begin{itemize}[leftmargin=*] 
\compactlist
	\item Can understand the sentiments and feelings.
	\item Can analyze the users' speech input by using linguistic analysis to 			detect emotion and language tones.
\end{itemize} \nointerlineskip  & 
\begin{itemize}[leftmargin=*] 
\compactlist
	\item Can convey feelings when responding to queries; such as anger, comfort 		and etc.
\end{itemize} \nointerlineskip  & 
\begin{itemize}[leftmargin=*]
\compactlist 
	\item ToneAnalyzer
	\item VoiceActivityDetect- ion
\end{itemize} \nointerlineskip \\ \bottomrule

\end{tabular}
\end{table*}

\subsection{Application of framework}

Our proposed framework has several use-cases, depending on the perspective of the stakeholder.
Platform users can use the framework for informing their investment into particular platforms, a decision that might require to carefully trade off conversational maturity with orthogonal factors such as cost and required resources.
Developers can use the framework as a benchmark to evaluate and compare their own platforms, to identify missing functionalities, and to make an informed decision on the "next steps" when extending their platform or developing a new platform from scratch.
Researchers can use the framework as a tool for obtaining a qualified overview of the platform landscape in practice.
In what follows, we describe the results of applying the framework to the twelve platforms considered in our survey.

Table \ref{tbl:applied}
 shows an overview of the analyzed platforms together with their proposed conversational maturity level, based on a manual assessment by the authors.
This assessment was done based on a feature matrix derived from the literature survey (see our online appendix \cite{onlineappendix}).
For each platform, we used the information on included features to map the platform to the corresponding level.
The table also includes a hint for features required to achieve the next level, which can act as a suggestion for current developers of the platforms to specifically 
focus on during development.
This is also a way to indicate for the
platform company on what features are necessary for the platform to
reach a more human-like performance if this is the end goal of the
platform.

The decision whether a platform has a particular feature or not is binary. 
One might further be interested in \textit{how well} each platform supports each feature.
To this end, one might combine our framework with a metrics-based approach such as the one by Venkatesh et al. \cite{Venkatesh}.
Having an automated assessment tool, ideally based on a continuously updated "benchmarking bot" supported by a community consensus, is a desirable direction for future work.

\section{Threats to validity}

The main threat to external validity is the question of how our framework will generalize to future platforms:
Since our framework based on an analysis of the documentation of existing platforms, it is essentially an overview of ``what we have'', rather than ``what we want''. 
A promising direction for future work is to conduct a user-study to identify the limitations and unaddressed user needs in current platforms.
Consequently, the framework might be augmented by a fifth maturity level.
Nevertheless, this concern does not threaten the usefulness of our framework for assessing current and near-future platforms and guiding organizations and developers in systematically improving their platforms.


A threat to internal validity are possible inaccuracies resulting from our decision to consult the considered platforms' documentations:
documentations might be incomplete (i.e., not all features are documented), and/or affected by overselling (i.e., documented features are actually not available).
While we cannot rule out that our findings are affected by this threat, we point out that companies have a significant interest in keeping their documentation up-to-date and complete.
Accurately depicting all available features is required to be competitive in a rising market; conversely, overselling might affect reputation and customer trust.
Furthermore, one may argue that our features and levels were defined in a subjective fashion (albeit with agreement between the authors); a large-scale user-rating study might increase objectivity.

\section{Conclusion}
We presented a conversational AI maturity framework for assessing conversational AI development platforms, based on the ability of the produced systems to conduct conversations. By supporting the understanding of how the features of a conversational AI development platform correspond to conversational ability, this framework can help both users with choosing and developers with developing a powerful conversational AI system. Our framework is inspired by related frameworks for human language development.
Comparable to the way in which a human speaker learns a language, the levels of conversational maturity in our framework indicate the ability to conduct and engage in a natural conversation with a user.

Our framework is based on, and incorporates results from an analysis of the state-of-the-art conversational AI development platforms, which we identified in a literature review.
We considered the documentations of these platforms to extract common and unique features, which we grouped into a feature model to provide a high-level overview of all the different existing features.
Each feature comes with a description to support the understanding of its use, context, and scope.
We related the features to conversational maturity and used them to develop our framework's maturity levels.

Our results show that the various existing conversational AI development platforms share significant commonalities.
In the future, to bridge different terminologies and support users in flexibility choosing a platform according to their current needs, we aim to develop a domain-specific language together with code generators for the various platform.
Such an infrastructure allows developing a system on a high level, and transforming the specification into an implementation for a concrete platform.
It can also support the migration between different platforms when a platform with higher conversational maturity more becomes available.


\balance{}

\bibliographystyle{SIGCHI-Reference-Format}
\bibliography{doc}

\end{document}